# Quantization of Keplerian systems

Ari Lehto

*Helsinki University of Technology*
*Laboratory of Materials Science*
*P.O. Box 6200, FIN-02015 TKK*

**Abstract**
A mathematical model is given for the occurrence of preferred orbits and orbital velocities in a Keplerian system. The result can be extended into energies and other properties of physical systems. The values given by the model fit closely with observations if the Planck scale is chosen as origin and the process considered as volumetric doubling in 3- and 4-dimensions. Examples of possible period tripling are also given. Comparison is made with the properties of the basic elementary particles, the Solar system and other physical phenomena.

## 1 INTRODUCTION

Connection between the Planck scale and the real world has been long sought for. The problem is in the extreme values of the Planck scale physical quantities. The Planck energy $E_o=10^{22}$ MeV is far too large for any elementary particle, the Planck length $l_o=10^{-35}$ m correspondingly far too small for any real object, and the Planck period $\tau_o=10^{-43}$ s is extremely short for any real world event.

A process generating subharmonics instead of higher harmonics would decrease the Planck energy in a natural way by bringing about lower energy levels. One such a process is period doubling, a common property of non-linear dynamical systems. This type of a process creates a series of doubling periods and correspondingly halving frequencies and energies according to $E=h/\tau$. A series of doubling lengths ($l=c\tau$) is borne, too. Period tripling, quadrupling etc. are also general properties of nonlinear dynamical systems.

It has been shown (Lehto 1990) that several stationary properties of matter coincide with values calculated from the Planck scale by 3- and 4-dimensional volumetric doubling, a form of period doubling. The perceived values of physical quantities seem to be cube roots of the corresponding volumes, save the electric charge squared (proportional to electrostatic energy), which is a fourth root.

In this article we examine properties of a non-linear system in terms of *period* $\tau$ rather than continuous time *t*. The best known example of such a formulation is Kepler's law $r^3=a\tau^2$, where *r* is the radius of the orbit, *a* a constant and $\tau$ the period of revolution.

We formulate a differential equation for a test object under a central force field by utilizing Kepler's law and give a numerical solution demonstrating the volumetric doubling and quantization of orbital velocity. The preferred orbits thus obtained can also be expressed in terms of energy, magnetic moment and temperature. The elementary electric charge is obtained from the Planck charge by corresponding volumetric doubling in four dimensions. The theoretical results are compared with observations.

## 2 THEORY AND DEFINITIONS

### 2.1 Preferred periods in a Keplerian system

Kepler's law tells that $r^3 = \tau^2$ (omitting the constant of proportionality). Solving for $r$ and taking the second derivative with respect to period $\tau$ one obtains

$$\frac{d^2 r}{d\tau^2} = -a\tau^{-\frac{4}{3}} \quad (1)$$

where $a$ is a constant. By again applying Kepler's law, Eq. (1) can be written in the form

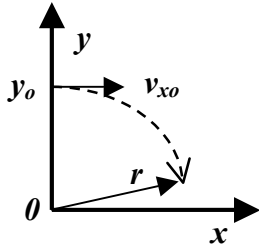

FIG.1. Initial setting of the test object.

$$\frac{d^2 r}{d\tau^2} = -\frac{a}{\tau^2} r \quad (2)$$

Equation (2) is formally equation of motion in terms of period of an oscillator, whose spring constant is inversely proportional to period squared, i.e. $a/\tau^2$. This means that the oscillations slow down with increasing period.

Direct solution of Eq. (2) would result in negative values of $r$ and therefore we rewrite Eq. (2) as follows:

$$\frac{d^2 x}{d\tau^2} = -\frac{ax}{\tau^2} \quad (3)$$

$$\frac{d^2 y}{d\tau^2} = -\frac{ay}{\tau^2} \quad (4)$$

The necessary requirement for circular motion is 90° phase difference between the $x$- and $y$-components. We shall now proceed with solving equations (3) and (4) simultaneously for a setting shown in figure 1 and reconstruct the radius using the Pythagorean theorem $r_z = sqrt(x^2 + y^2)$. The test object is initially located at $(0, y_o = 1)$ and given an initial velocity $(v_{xo}, v_{yo}=0)$. Equations are solved starting from initial period $\tau = 1$ and $a = 46.47014$. The initial $x$-component of velocity is $v_{xo} = sqrt(a)$.

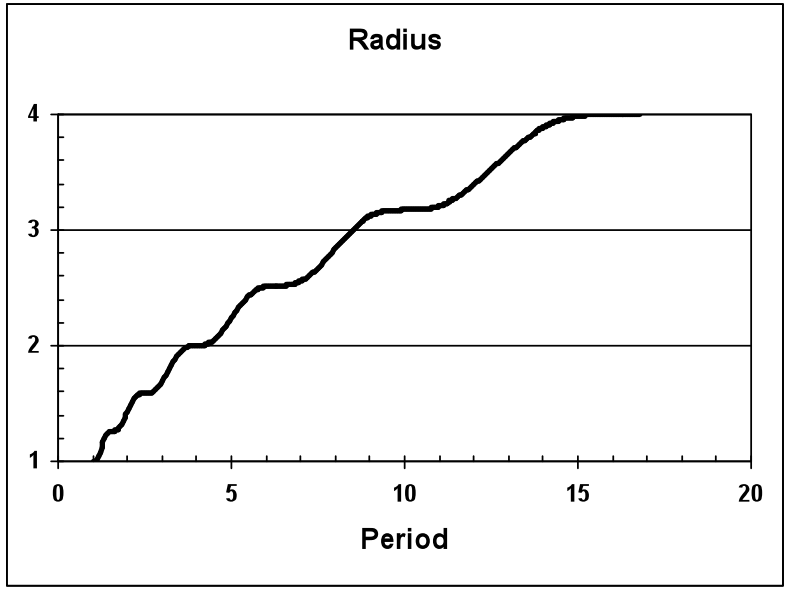

Figure 2 is a plot of a simultaneous numerical solution to Eq's (3) and (4) showing radius $r$ as function of period $\tau$. One can see that there are plateaus, where radius $r$ remains constant over a range of periods. The ratio of adjacent radii (plateau values) is cube root of two.

FIG. 2. Radius as function of period showing ranges of period with constant radii.

Figure 3 shows volume $r_z^3$ as function of period. With the parameters given the plateaus occur at doubling volumes (1) 2, 4, 8, 16, 32 etc.

Corresponding calculation for $r_x$ and $r_y$ can be carried out (rotation about $x$- and $y$-axes). Here we limit ourselves to identical rotations about the three axes and the volume is therefore represented by $V=r_z^3$ (cube) rather than by $V=r_x r_y r_z$ (parallelepiped). We could have also written the volume of the 3-dimensional structure in terms of period, angular frequency or energy.

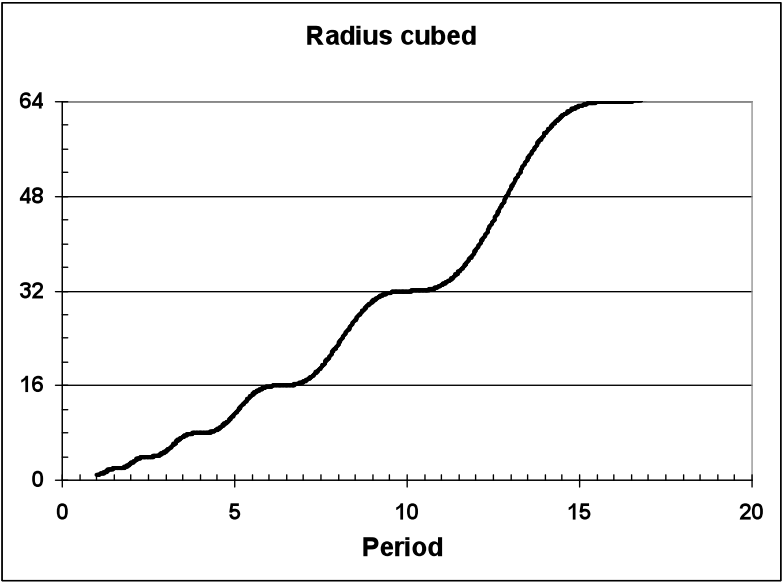

FIG. 3. Plateaus in volume $r_z^3$ as function of period.

Figure 4 shows the orbital velocity $v=2\pi r_z/\tau$ of the test object as function of the radius. It can be seen that the velocity of the test object is quantized and the ratio of adjacent values is cube root of two.

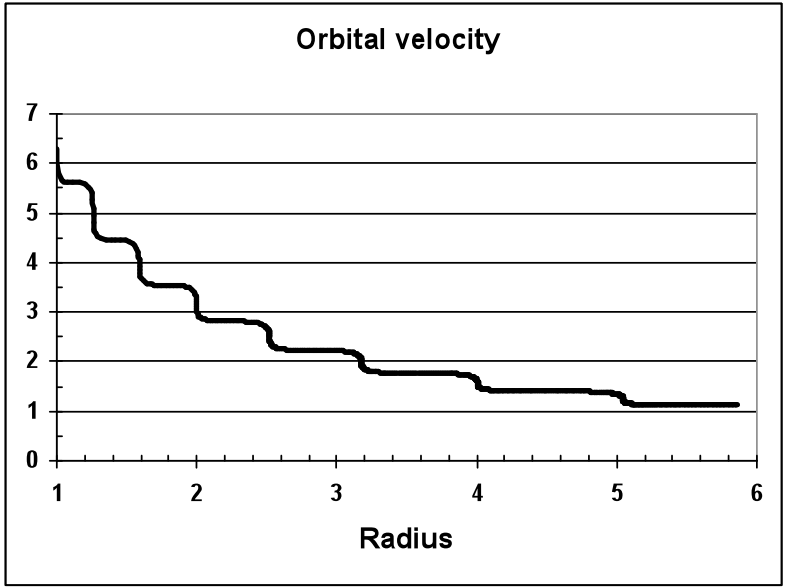

FIG. 4. Orbital velocity of the test object as function of the orbital radius.

## 2.2 4-dimensional doubling

In principle, volumetric doubling in 4-dimensions does not differ from the 3-dimensional doubling. In this case the value of constant *a* is *a=82.4*.

## 2.3 The Planck scale units

The doubling process yields absolute and unadjustable values if the Planck scale is chosen as origin. We shall now define the Planck scale needed in the comparison of the values given by the model with observations. The necessary Planck scale units are defined as:

Period $\quad\quad\quad\quad\quad\quad\quad\quad\quad \tau_o = \sqrt{\dfrac{hG}{c^5}}$ (5)

Length $\quad\quad\quad\quad\quad\quad\quad\quad\quad l_o = c\tau_o$ (6)

Energy $\quad\quad\quad\quad\quad\quad\quad\quad\quad E_o = \dfrac{h}{\tau_o}$ (7)

Charge $\quad\quad\quad\quad\quad\quad\quad\quad\quad q_o = \sqrt{4\pi\varepsilon_o hc}$ (8)

Temperature $\quad\quad\quad\quad\quad\quad\quad\quad T_o = \dfrac{E_o}{k}$ (9)

Velocity $\quad\quad\quad\quad\quad\quad\quad\quad\quad c$ (10)

Units using the elementary electric charge are:

Coulomb energy $\quad\quad\quad\quad\quad E_o^E = \dfrac{1}{4\pi\varepsilon_o}\dfrac{e^2}{l_0}$ (11)

Magnetic moment $\quad\quad\quad\quad\quad \mu_{oorb} = \dfrac{ec^2}{4\pi}\tau_o$ (12)

Magnetic moment $\quad\quad\quad\quad\quad \mu_{orad} = \dfrac{\pi ec^2}{16}\tau_o$ (13)

Table I shows the values of the Planck scale units defined by equations (5) to (10) needed in this study.

Equation (7) is the Planck relation *E=hf*, where *f=1/τ_o*. Equation (8) in turn defines the Planck charge, which is analogous to the Planck mass.

Equations (12) and (13) represent two different geometries, where the Planck length is either the circumference or the diameter of a magnetic moment current loop, respectively.

TABLE I. The Planck scale units.

|  | Symbol | Value | Unit |
|---|---|---|---|
| Period | $\tau_o$ | $1.3514 \cdot 10^{-43}$ | s |
| Length | $l_o$ | $4.0513 \cdot 10^{-35}$ | m |
| Energy | $E_o$ | $3.0603 \cdot 10^{22}$ | MeV |
| Charge | $q_o$ | $4.7013 \cdot 10^{-18}$ | As |
| Temperature | $T_o$ | $3.55 \cdot 10^{22}$ | K |
| Velocity | $c$ | 299792458 | m/s |

Table II shows unit values for the electrostatic energy and magnetic moment. These units are not strictly Planck scale units, because the elementary electric charge is used instead of the Planck charge. These units must be used because the particles do not carry the Planck charge but the elementary electric charge. It is shown in chapter 3.11.1 how the elementary electric charge is borne by the doubling process in four dimensions.

In this model part of the rest energy of charged particles is in their electric fields. The electrostatic (Coulomb) energy can be calculated from equation (11).

TABLE II. Auxiliary units.

|  | Symbol | Value | Unit | Comment |
|---|---|---|---|---|
| Coulomb energy | $E_o^E$ | $3.5543 \cdot 10^{19}$ | MeV | $(e^2/q_o^2) \cdot E_o$ |
| Magnetic moment | $\mu_{o\,orb}$ | $1.5485 \cdot 10^{-46}$ | Am$^2$ | Orbital quantization |
| Magnetic moment | $\mu_{o\,rad}$ | $3.8208 \cdot 10^{-46}$ | Am$^2$ | Radial quantization |

The gravitational constant $G$ is the main source of inaccuracy in the Planck scale units. The relative standard uncertainty of $G$ is 150 ppm and hence that of the Planck energy, length, and temperature 75 ppm. The values of the natural constants have been adopted from the NIST (Mohr and Taylor 2005) table of natural constants.

## 2.4 Notation

In a three dimensional Cartesian coordinate system volume is $V_N = 2^i \cdot l_o \cdot 2^j \cdot l_o \cdot 2^k \cdot l_o = 2^{i+j+k} \cdot l_o^3 = 2^N \cdot V_o$, where $V_N$ is the volume of the object, $i, j, k$ numbers of doubling of the edge lengths and $V_o$ the initial or unit volume (e.g. Planck scale cube). Corresponding "volumes" can be written for any physical quantity derivable from the Planck length.

The total number of doublings $N=i+j+k$ will be denoted by $(i, j, k)$, which also refers to the structure of the object. The corresponding notation in 4-d is $(i, j, k, l)$. The perceived number of doublings is denoted by $n=N/3=(i+j+k)/3$ and $n=N/4=(i+j+k+l)/4$ for 3- and 4-dimensions correspondingly (cube and fourth roots).

The geometric shape of the object or structure defined by its edges is normally parallelepiped but the shape is cubical, if $i, j, k$ or $i, j, k, l$ are all equal.

The charge squared ($q_n^2$) with $n=9.75$ is called the elementary electric charge (squared) and denoted by $e^2$.

Magnetic moment will be denoted by $\mu$ and defined classically as current times loop area, or $\mu=iA$.

The rest energy of the electron can be related to a single Planck energy sublevel. This kind of a particle will be called single-level particle. The corresponding Planck energy is denoted by $E_o$. Particles, whose rest energy is determined by the sum energy of two sublevels, will be called sum-level particles. The corresponding Planck energy is denoted by $E_{os}$. Sublevel means any energy level obtained from the Planck energy by the frequency ($f=1/\tau$) halving or period doubling process.

## 2.5 Basic formulae

The preferred orbits enable us to derive formulae for physical quantities corresponding to the quantization. Preferred energies can be calculated by using the Planck relation $E=hf=hc/l_o$ and temperatures by $T=E/k$ ($k$=Boltzmann's constant). Magnetic moment can be calculated from $\mu=iA$ (current loop) by the 3-dimensional doubling process and electric charge from the Planck charge $q=sqrt(4\pi\varepsilon_o hc)$ by the four dimensional inverse doubling (i.e. halving, negative exponent) process. Because energy is proportional to charge squared, we calculate the subcharges using the Planck charge squared. The basic formulae are:

Energy level (3-d) $\quad E_n = 2^{-N/3} \cdot E_o = 2^{-n} \cdot E_o \quad$ (14)

Energy level (4-d) $\quad E_n = 2^{-N/4} \cdot E_o = 2^{-n} \cdot E_o \quad$ (15)

Charge squared (4-d) $\quad q^2_n = 2^{-N/4} \cdot q_o^2 = 2^{-n} \cdot q_o^2 \quad$ (16)

Magnetic moment (3-d) $\quad \mu_n = 2^{N/3} \cdot \mu_o = 2^n \cdot \mu_o \quad$ (17)

Length (3-d) $\quad l_n = 2^{N/3} \cdot l_o = 2^n \cdot l_o \quad$ (18)

Period (3-d) $\quad \tau_n = 2^{N/3} \cdot \tau_o = 2^n \cdot \tau_o \quad$ (19)

Temperature (3-d) $\quad T_n = 2^{-N/3} \cdot T_o = 2^{-n} \cdot T_o \quad$ (20)

Velocity (3-d) $\quad v_n = 2^{-N/3} \cdot c = 2^{-n} \cdot c \quad$ (21)

Equation (21) can be derived as follows: $v_{ij}=l_i/\tau_j=2^i l_o/2^j \tau_o=2^{i-j}l_o/\tau_o=2^{i-j}c=2^{-n}c$, where $c$ is the speed of light and $l_o/\tau_o=c$ according to equation (6).

Equations (14) to (21) can be written as one constitutive equation:

$$X_n = 2^{\pm n} \cdot X_o \quad (22)$$

where $X_o$ is any Planck scale unit.

## 2.6 Separation of adjacent levels

In the three dimensional system the adjacent values $X_i=2^i X_o$ and $X_j=2^j X_o$ ($j=i+0.333$) are separated by a factor of $2^{0.333}$, which means that $(X_j-X_i)/X_i=260000$ ppm and $(X_i-X_j)/X_i=210000$ ppm. The separation is correspondingly 190000 ppm and 160000 ppm in four dimensions. $X$ is any quantity in equations (14)-(21). Period tripling separates levels by $3^{0.333}$, or about 400000 ppm (40%). For the accuracy of the model the difference between the calculated and experimental values may be compared with the level separations.

## 2.7 Density of states

The perceived density of the Planck energy sublevels or states, $D(E)=\Delta n/\Delta E_n$, can be calculated from equations.(14) and (15) by solving for $n$. The (absolute) density of states is

$$D(E) = \frac{i}{E_n \ln 2} = \frac{i}{E_o \ln 2} 2^n \quad (23)$$

where $i$ is 3 or 4 depending on the number of dimensions. Equation (23) shows that the density of states grows exponentially with $n$. The perceived Planck scale 3-d density of states is $D(E)=1.4\cdot10^{-28}$ (1/eV).

## 2.8 Superstability

The result of a doubling process is a quantized system with exact values. Considering transitions it is customary to talk about initial and final states. From the operational viewpoint transition to a new state results from an operation on the initial state.

The $1/x$ shape of both the Coulomb and gravitational potentials leads to $x^3=a\tau^2$ dependence of $x$ on $\tau$ (Keplers's law). Let us now define volume $V$ as $V=x^3$ and consider (the expanding) $V$ as driving force or an operator $V$ acting on the period $\tau$. According to $V=\tau^2$ the operation is squaring the period. Let us further assume that there is a shortest period $\tau_o$, which doubles (as space expands) according to $\tau/\tau_o=2^i$, where $i$ is an integer.

The first operation of $V$ on $\tau/\tau_o$ yields $V\mathrm{o}(\tau/\tau_o)=2^1$ (where o means operation), the second operation is $V\mathrm{o}(V\mathrm{o}(\tau/\tau_o))=(2^1)^2=2^2$, the third $V\mathrm{o}[V\mathrm{o}(V\mathrm{o}(\tau/\tau_o))]=(2^2)^2=2^4$, the fourth $V\mathrm{o}\{V\mathrm{o}[V\mathrm{o}(V\mathrm{o}\,(\tau/\tau_o))]\}=(2^4)^2=2^8$ and so forth. The resulting periods can be represented as

$$\tau_i = 2^{2^i}\tau_o \qquad (24)$$

where $i$ is an integer. This is the result of functional iteration leading to superstable periods, as shown by Feigenbaum (1980). The total number of doublings is of the form $N=2^i$.

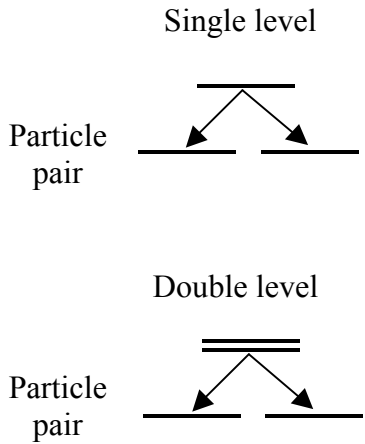

FIG. 5. Production of particle pairs.

## 2.9 Particle creation

Electron-positron pair creation from a 1.022 MeV gamma quantum is perhaps the best known example of materialization of energy. We assume that this type of a pair conversion process is generally valid and applicable to the Planck energy sublevels obtained by the volumetric halving process. The particle production principle is illustrated in figure 5. A single sublevel and a double level split into a pair of particles.

Double level is either a sum or difference of two sublevels. For the nucleons we consider the double level as a sum level.

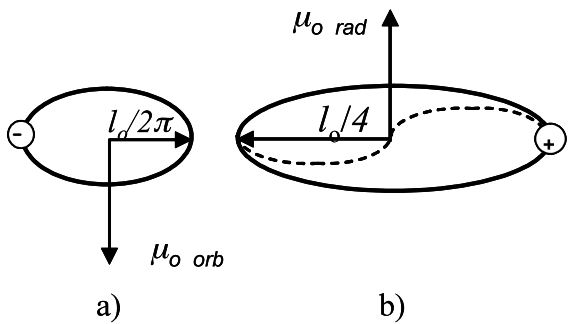

FIG. 6. Current loops defining the two Planck scale magnetic moments, a) for the orbital geometry and b) for the radial geometry.

## 2.10 Magnetic moment

Let us now define the unit magnetic moment $\mu_o$ as a classical current loop in the Planck scale. By definition magnetic moment equals current times the loop area. The loop current is obtained by dividing the elementary charge $e$ by the period of orbital revolution. Two different loops, shown in figure 6, are defined: a) the orbital type denoted by $\mu_{o\ orb}$ with the Planck length $l_o$ as the circumference, and b) the radial type denoted by $\mu_{o\ rad}$ with half of the Planck length as the diameter of the loop. This geometry corresponds to a potential well ground state, where the width of the well is one half of the wavelength. The definition of the orbital type unit magnetic moment $\mu_{o\ orb}$ (for e.g. the electron) is:

$$\mu_{oorb} = \frac{e}{\tau_o}\pi(\frac{l_o}{2\pi})^2 = \frac{ec^2}{4\pi^2}\cdot\tau_o \quad (25)$$

The unit radial type magnetic moment $\mu_{o\ rad}$ (for e.g. the nucleons) is correspondingly:

$$\mu_{orad} = \frac{e}{\tau_o}\pi(\frac{l_o}{4})^2 = \pi ec^2\cdot\frac{\tau_o}{16} \quad (26)$$

The numeric value of $\mu_{o\ orb}$ = 1.5485·10$^{-46}$ (Am$^2$) and $\mu_{o\ rad}$ = 3.8208·10$^{-46}$ (Am$^2$).
The direction of the moment is perpendicular to the plane of the loop as shown in figure 6. If particles are considered as four dimensional objects in a Cartesian coordinate system, then the magnetic moment vector can be oriented to any direction. The zero magnetic moment of the mesons suggests that their magnetic moment is directed to the fourth dimension and therefore there is no vector component in the other three dimensions (i.e. in our 3-d space).

## 3 EXPERIMENTAL
## 3.11 Elementary particles

### 3.11.1 Elementary electric charge and the fine structure constant

The Planck charge $q_o$=4.701·10$^{-18}$ (As) is surprisingly close to the elementary electric charge $e$=1.602·10$^{-19}$ (As) differing only by a factor of about 29. Obviously equality would be the simplest case, but it seems that subcharges have been borne in process of time, as doubling (i.e. halving) process has continued. We shall now show that a particular subcharge is the elementary electric charge.
The perceived number of doublings for the elementary charge squared is:

$$n = \log(\frac{e^2}{q_o^2})/\log(2) = -9.7499 = -\frac{39}{4} \quad (27)$$

which means $N$=39 doublings in four dimensions. According to equation (16) a perceived charge $e^2$ is created:

$$e^2 = g\cdot q_o^2 \quad (28)$$

where $g$ is $2^{-39/4} = 2^{-(1+2+4+32)/4}$, which is a superstable (1, 2, 4, 32) structure.

The electric force constant is called the fine structure constant alpha and defined as

$$\alpha = \frac{e^2}{2\varepsilon_o hc} \quad (29)$$

By dividing both sides by $2\pi$, one obtains $\alpha/2\pi=e^2/4\pi\varepsilon_o hc$ or

$$\frac{\alpha}{2\pi} = \frac{e^2}{q_o^2} \quad (30)$$

which is the ratio of the elementary charge and the Planck charge squared. We further obtain

$$\alpha = 2\pi \frac{e^2}{q_o^2}$$

Since $e^2/q_o^2 = 2^{-39/4}$, the inverse of the fine structure constant $\alpha$ obtains a value of $\alpha^{-1} = 2^{39/4}/2\pi = 137.045$. The difference to the recommended value is 65 ppm.

The value of the elementary charge is obtained from Eq. (28):

$$e = 1.60213 \cdot 10^{-19} \text{ (As)}, \quad (31)$$

which differs by *30* ppm from the recommended value.
The process of volume doubling may also produce other values for the force constants. Some of these have been calculated (Lehto 1990).

### 3.12 Rest energy and magnetic moment

#### 3.12.1 Charged leptons

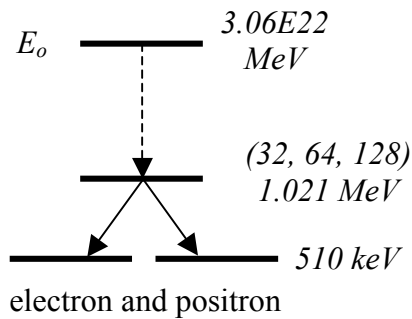

FIG. 7. Planck sublevels for the electron-positron pair.

Figure 7 shows the superstable $N=224$ or (32, 64, 128) sublevel energy of 1.021 MeV, which is almost exactly the same as the rest energy of an electron-positron pair. We may therefore assume that under proper conditions this sublevel may materialize into an electron-positron pair, when split into two equal parts and occupied by opposite elementary electric charges. The total energy of the object is the sublevel Planck energy added by the corresponding electric field Coulomb energy.

Equations (25) and (26) show that magnetic moment is directly proportional to period. Energy ($E=h/\tau$), in turn, is inversely proportional to period. The splitting of the $N=224$ sublevel energies and magnetic moments then corresponds to $N=227$ halvings and doublings and $n=75.67$ applies for both the half-energies (0.510 MeV) and the magnetic moments, as shown in tables III and IV.

TABLE III. Planck level $N=224$ energy (MeV).

|  | Energy | $n$ | $N$ | Comment |
|---|---|---|---|---|
| Planck energy | 1.0206 | 74.67 | 224 | (32, 64, 128) |
| Coulomb energy | 0.0012 | 74.67 | 224 | (32, 64, 128) |
| Total energy | 1.0218 |  |  |  |
| Half energy | 0.5109 | 75.67 | 227 | Electron |

TABLE IV. Level $N=224$ orbital type magnetic moment.

|  | $\mu$ | $n$ | $N$ | Comment |
|---|---|---|---|---|
| Calculated | $4.643 \cdot 10^{-24}$ | 74.67 | 224 | (32, 64, 128) |
| 2x calculated | $9.286 \cdot 10^{-24}$ | 75.67 | 227 | Doubling |
| Electron | $9.285 \cdot 10^{-24}$ |  |  | Experimental |

The model for an electron-positron pair is simply the Planck 1.021 MeV sublevel occupied by opposite elementary electric charges. The differences between the calculated and experimental values may be compared to the separation of the theoretical 3-d sublevels in chapter 2.6 showing that the measured values are very close to the calculated ones.

Table V shows the structures of the electron-positron pair and the muon. The muon rest energy 105.6 MeV is close to the uncharged $n=68$ Planck level ($E=103.7$ MeV).

The rest energy of the heavy lepton $\tau$ (1777) does not correspond to any single Planck level. However, the energy difference between $E_{os}$ series level $n=63.75$ and $E_o$ series level $n=63$ is 1776.3 MeV.

TABLE V. Electron-positron pair and the muon structures.

| Particle | $n$ | $N=4n$ | Components | | | |
|---|---|---|---|---|---|---|
|  |  |  | $i$ | $j$ | $k$ | $l$ |
| $e^-+e^+$ | 74.67 | 224 | 32 | 64 | 128 | 0 |
| muon | 68.00 | 272 | 64 | 64 | 128 | 16 |

Table V also shows that the muon components can be transformed into the electron-positron pair components by a three dimensional doubling, where $i=64$ becomes $i=32$. The $N=224$ or 1.021 MeV level can further split into two 0.511 MeV levels. Because there is only one elementary charge available the other 0.511 MeV level remains unoccupied and only an electron appears.

**Electron magnetic moment anomaly**

The magnetic moment $\mu$ of a particle is traditionally considered as resulting from the charge $e$, mass $m$ and an internal spin $S$ ($= h/2\pi$) of the particle:

$$\mu = g_s \cdot \frac{e}{2m} \cdot S \qquad (32)$$

where $eS/2m$ is called a magneton (either Bohr or nuclear depending on the value of $m$). The dimensionless number $g_s$ is (experimental) gyromagnetic ratio thought to reflect the internal structure of the particle. The values of $g_s$ for the electron, proton and neutron are $-1.001$, $+2.793$ and $-1.913$ respectively. The measured value of the electron magnetic moment is a little larger than the Bohr magneton $\mu_B=eS/2m_e$. Magnetic moment anomaly is defined as $a_e = |\mu_e|/\mu_B - 1$, where $|\mu_e|$ is the measured electron magnetic moment. The anomaly can be measured very accurately and its current accepted value is $a_e = 0.00116$.

In this model the magnetic moment anomaly is obtained by replacing the Bohr magneton by the magnetic moment value given in table IV for the electron. One obtains

$$a_e = \frac{|\mu_e|}{\mu_{model}} - 1 = -0.00016 \qquad (33)$$

which is negative and about one tenth of the classical anomaly. The value of $a_e$ differs by 160 ppm from unity. This discrepancy may be partly due to the inaccuracy of $G$ in the Planck mass and hence in the unit magnetic moment. In principle the doubling process is exact.

### 3.12.2 Neutrino

The Planck energy level system can be considered as a multi-dimensional lattice, because the energy values are fixed and a geometric structure or a shape can be assigned to. The role of a neutrino in particle processes within this lattice may be considered as analogous to the role of phonons in electronic transitions in crystal lattices, namely the transfer of momentum to the surrounding lattice as a whole.

### 3.12.3 Mesons

Table VI shows the rest energies of some meson pairs as compared to the corresponding Planck sublevel energies. An overall agreement is found.

TABLE VI. Meson energies and number of doublings.

|  | Measured energy | Planck energy | $E_o$ $n$ | $E_{os}$ $n$ |
|---|---|---|---|---|
| $\pi^+ + \pi^-$ | 279 | 279 |  | 66.75 |
| $K^+ + K^-$ | 987 | 986 | 64.75 |  |
| $D^+ + D^-$ | 3739 | 3749 |  | 63.00 |
| $F^+ + F^-$ | 3937 | 3946 | 62.75 |  |
| $B^+ + B^-$ | 10559 | 10604 |  | 61.50 |

It is customary to consider pions and kaons as triplets and figure 8 shows that a Planck sublevel can be split into meson triplets. Energy level $N=4*66=264$ splits into pions and $N=3*65.33=196$ into kaons. Note that only the three dimensional part of the kaons is perceived, which makes a strange distinction between these particle triplets. This is pair production with a neutral particle.

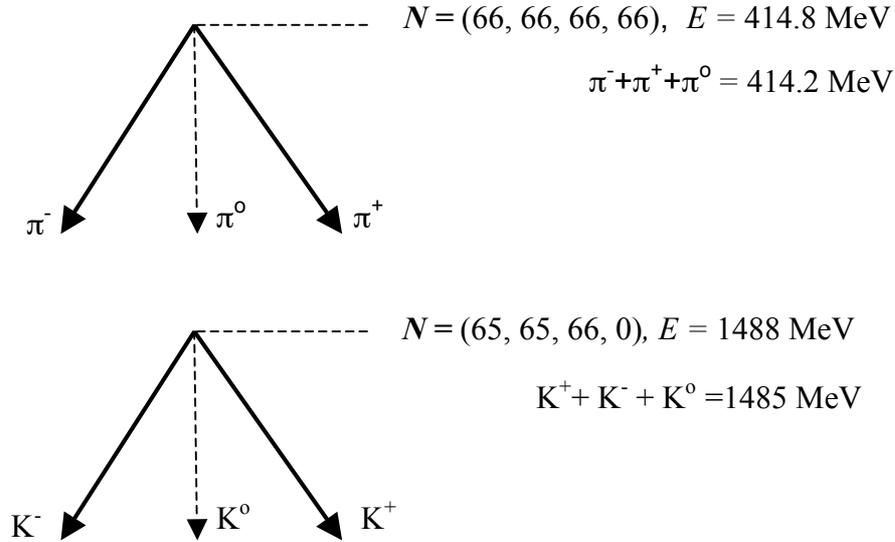

$N = (66, 66, 66, 66)$, $E = 414.8$ MeV

$\pi^- + \pi^+ + \pi^o = 414.2$ MeV

$N = (65, 65, 66, 0)$, $E = 1488$ MeV

$K^+ + K^- + K^o = 1485$ MeV

FIG. 8. Pion and kaon triplets and corresponding total number of doublings. The kaon triplet is perceived as a three dimensional object.

### 3.12.4 Pion decay

The charged pions decay into muons. Table VII shows that in the decay process the total number $N=4n$ of doublings increases by one (from 271 to 272) or $l$ from 15=16-1 to 16.

TABLE VII. Charged pion decay into a muon.

| Particle | n | N=4n | Components | | | |
|---|---|---|---|---|---|---|
| | | | i | j | k | l |
| pion | 67.75 | 271 | 64 | 64 | 128 | 15 |
| muon | 68.00 | 272 | 64 | 64 | 128 | 16 |

### 3.12.5 Nucleons

The leptons and baryons differ from one another in that the leptons are considered as structureless pointlike particles, whereas the baryons have measurable size and some kind of an internal structure. This difference is also reflected in the way their rest energies and magnetic moments are borne in the doubling process.

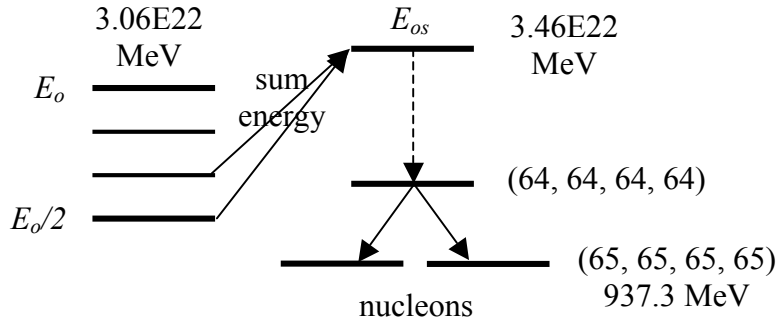

FIG. 9. Planck sublevels for the nucleons from sum energy.

The nucleon pair can be generated from sum energy level $E_{os}$ as shown in figure 9. The (64, 64, 64, 64) sublevel with $2^{-64} \cdot E_{os} = 1874.6$ MeV energy splits into two 937.3 MeV energy levels. The (64, 64, 64, 64)-level magnetic moment correspondingly doubles into two 65-level magnetic moments.

TABLE VIII. Planck level $n=64$ or $N=256$ energy (MeV).

| n | N | proton | neutron | Comment |
|---|---|---|---|---|
| 64.00 | 256 | 1874.6 | | N=4n |
| 65.00 | 260 | 937.3 | 937.3 | nucleon Planck energy |
| 74.33 | 223 | | 1.286 | N=3n |
| 74.66 | 224 | 1.021 | 1.021 | N=3n |
| | sum | 938.32 | 939.61 | calculated |
| | measured | 938.28 | 939.57 | proton and neutron |

The experimental rest energies of the proton and neutron are 938.27 MeV and 939.57 MeV respectively. Table VIII shows that 937.31 MeV energy is a little short of the rest energies of the nucleons. If we attach a superstable (32, 64, 128) uncharged sublevel structure and energy to the 937.31 MeV level, a very close agreement between the calculated and measured proton rest energies is obtained.

The measured difference between the rest energies of the neutron and the proton is 1.28 MeV. Table VIII shows that this energy corresponds to $n=74.33$ or $N=223$ sublevel energy, adjacent to the superstable $N=224$ sublevel. The elementary electric charges can be attached to the 937.31 MeV structure by adding an electron-positron pair structure with one or two charges.

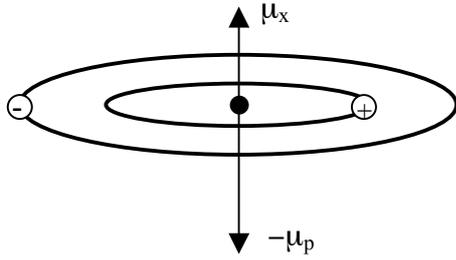

FIG. 10. Neutron's current loops.

The $N=192$ or (64, 64, 64, 64) radial type magnetic moment is $7.0482 \cdot 10^{-27}$ Am$^2$. If this value is multiplied by two (corresponding to (65, 65, 65, 65) structure obtained by doubling) the magnetic moment of a proton is obtained, as shown in table IX.

According to measurements neutron's charge is divided into a positively charged inner layer and a negatively charged outer layer. There are thus two current loops of opposite magnetic moments. The negative loop has larger magnetic moment than the positive one and the total (negative) magnetic moment is the sum of the moments of the two loops.

Simplest case is now assumed: The larger loop is identical to the proton's loop $\mu_p$ (save the sign of the charge) and a concentric smaller positive loop $\mu_x$ is added, as shown in figure 10.

The measured magnetic moment of the neutron is minus $9.6624 \cdot 10^{-27}$ (Am$^2$). The unknown $\mu_x$ can now be calculated from $\mu_x = \mu_p + \mu_n$. It is found that

$$\mu_x = 2^{63.33} \cdot \mu_{o\ rad} \tag{34}$$

or

$$\mu_x = 2^{-1.67} \cdot \mu_p$$

which shows that the magnitude of $\mu_x$ results from equation (17), too. Neutron's magnetic moment is then:

$$\mu_n = (2^{63.33} - 2^{65.00}) \mu_{o\ rad} \tag{35}$$

and no experimental gyromagnetic ratio is needed. Table IX shows the calculated and measured magnetic moments of the nucleons.

TABLE IX. Magnetic moments of the nucleons.

| $n$ | Nucleon | $\mu$ (Am$^2$) | Measured | Comment |
|---|---|---|---|---|
| 65.00 | proton | $1.4096 \cdot 10^{-26}$ | $1.4106 \cdot 10^{-26}$ | positive loop |
| 65.00 | neutron | $-1.4096 \cdot 10^{-26}$ | | negative loop |
| 63.33 | | $4.4399 \cdot 10^{-27}$ | | positive loop |
| | sum | $-9.656 \cdot 10^{-27}$ | $-9.662 \cdot 10^{-27}$ | |

The experimental magnetic moment of the proton differs from the calculated one by 720 ppm. The difference for the neutron is 660 ppm.

A sublevel is just an energetic state without any electrical charges. In principle it can be neutral or occupied by one charge or two (opposite) charges. In this model the neutron is neutral, because it hosts two elementary charges forming the magnetic moment current loops but there is only one positive charge within the proton.

As hydrogen is the most abundant substance in the universe, it is plausible to think that the negative charge associated with the proton's positive elementary charge remains attached to the proton, although rather loosely. The ground state orbital magnetic moment of hydrogen is the same as electron's (intrinsic) magnetic moment. This means that the Bohr orbit's magnetic moment has the same periodic structure as the electron's magnetic moment.

The special feature with the proton is that the same perceived number of doublings, i.e. $n=65.00$, is obtained from both a 3-d and a 4-d object ((65, 65, 65) and (65, 65, 65, 65)).

### 3.12.6 Baryon decay

Figure 11 shows the observed high probability baryon decays. It can be seen that the decay is related to structural component changes (right side of figure).

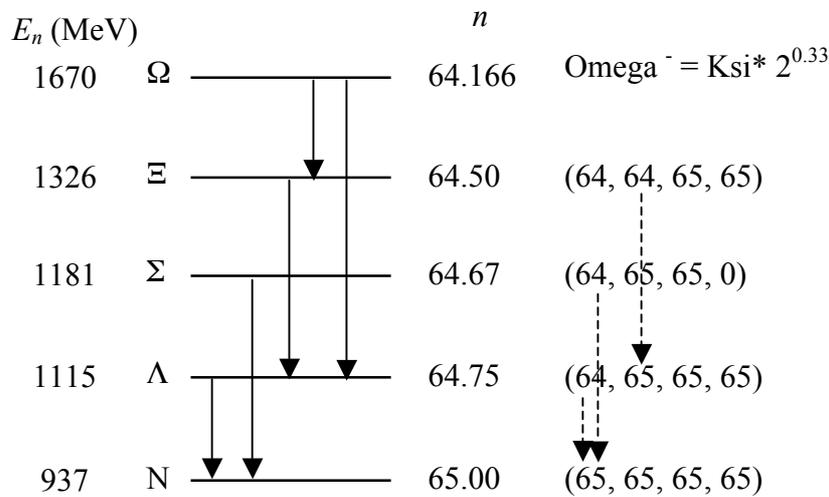

FIG. 11. Baryon decay schematically.

The perceived sigma particle is apparently 3-dimensional, so it decays into the 3-dimensional part of the nucleons.

## 3.13 Particle collisions

Figure 12 shows a standard example of proton-antiproton collision producing a ksi particle pair. The proton-antiproton pair energy corresponds to the Planck energy with $N=256$. Corresponding number of doublings for the ksi pair is $N=254$. This means that two components have changed by unity.

Figure 13 shows another standard example of proton-antiproton collision (cascade).

First a charged pion pair and an eta particle are borne followed by the immediate decay of the eta into a pion triplet.

It can be seen that the doubling takes place in all four dimensions at the corresponding Planck energy levels.

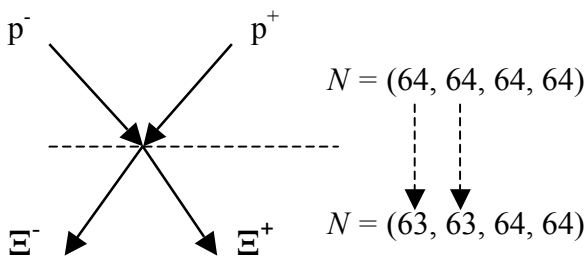

**FIG. X**. Proton-antiproton collision producing two ksi particles. Doubling occurs in two components.

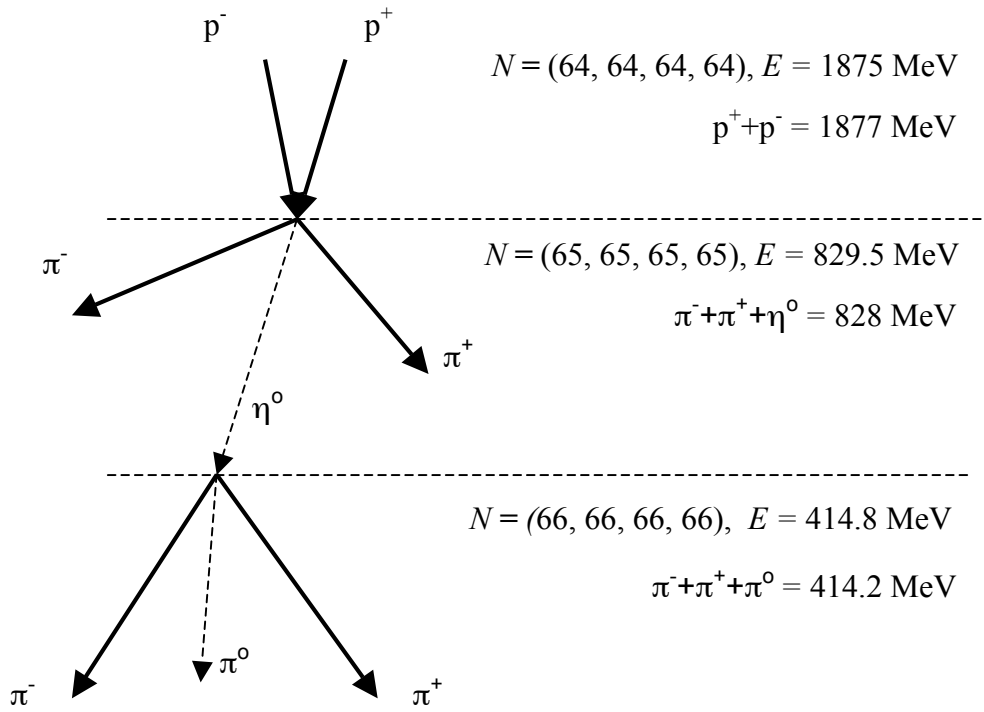

FIG. 13. Proton-antiproton collision cascade showing doubling in all four dimensions.

## 3.14 The Solar system

Equations (18) and (21) yield the model lengths and velocities. Figure 14 shows the distances and orbital velocities of the planets in (*r,v*)-space. The asteroids (*N*=42 in equation (21)) occupy the gap between Mars and Jupiter. The observed velocities are consequent values of *N* from *N*=38 (Mercury) to *N*=48 (Pluto) in equation (21) with one exception. There is an unoccupied orbit with *N*=44 between Jupiter and Saturn, as if a planet were missing there, too.

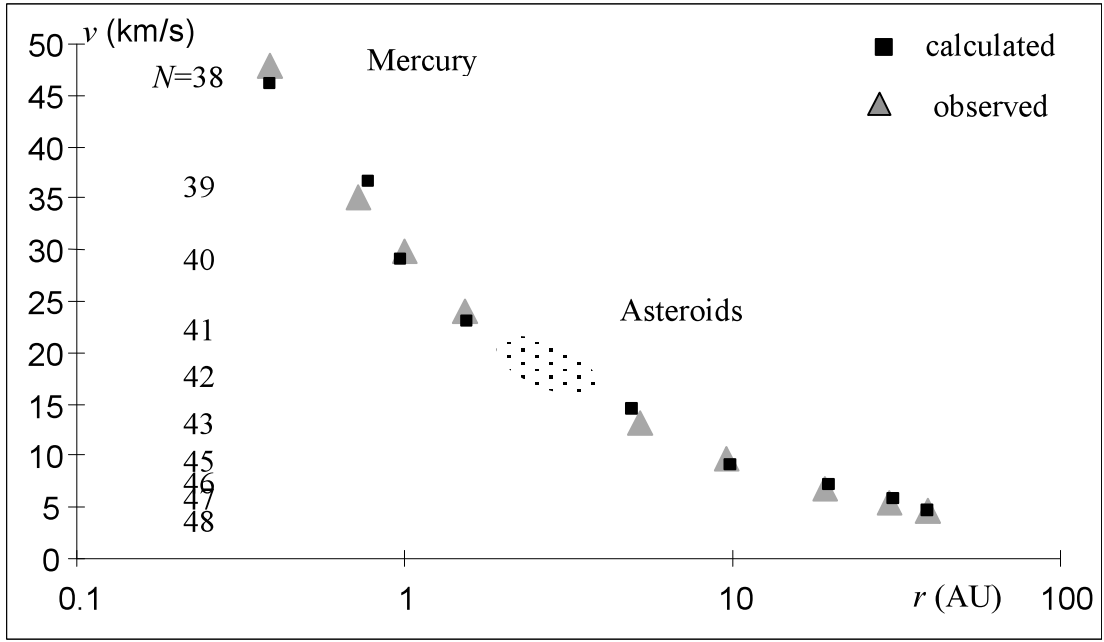

FIG. 14. Planets in $(r,v)$ space. The asteroids occupy space around $N=42$ orbit and $N=44$ is empty. Note that $N=42$ and $N=44$ are on both sides of Jupiter ($N=43$). The $N$-values indicated belong to $v$ in Eq. (21).

The distances from the Sun have been calculated from equation (18) with the Planck length as origin.

## 3.15 Quantized galaxy redshifts

According to the accepted cosmological model space is expanding. This view is based on the observed redshifts of the galaxies, which is the larger the dimmer (i.e. farther away) the galaxy is. This observation led to the idea that redshift is due to the Doppler shift, meaning that all galaxies recess from us the faster the farther away they are.

W.G. Tifft of the University of Arizona has measured redshifts for over twenty-five years using both optical and radio spectra (at 21 cm wavelength). Careful measurements with large radio telescopes have given results far exceeding the accuracy of optical measurements. Tifft (1997) has found out with high S/N ratio that the redshifts are not only quantized but also variable.

Quantization has been verified by Napier and Guthrie (1996) and Napier (2003). These observations indicate that the redshift is not due to motion alone.

The most prominent redshift period corresponds to 73 km/s and its half. Practically all Tifft´s redshift periods follow Eq. (21) and its modification (Lehto-Tifft rule), where cube root is replaced by ninth root (allowing transitions between redshift states). According to the cosmological principle all observers anywhere in the universe should observe the same redshift periods. The origin of the quantized redshift may not necessarily be velocity at all but an expression of the energetic state of the galaxy. If the volume of the universe is expanding, then doubling proceeds in process of time, which may explain the variable redshift periods observed by Tifft.

## 3.16 Hydrogen and Deuterium hyperfine structure - period tripling

The radio frequency electromagnetic radiation emitted by Hydrogen and Deuterium hyperfine transitions, also called "spin-flip radiation", can be used e.g. in studies of cosmic abundances and distributions of these gases. The energy released in the Hydrogen transition is *E*=5.9 μeV and appears as *f*=1420 MHz radio frequency known as the 21 cm Hydrogen line. Deuterium emits radio waves at an energy of *E*=1.36 μeV, or *f*=327 MHz, corresponding to 92 cm wavelength. Both emissions can be received and analyzed using radio telescopes and related equipment.

Equation (19) can be written in terms of frequency as

$$f_N = f_o \cdot 2^{-N/3} \qquad (36)$$

where $f_N$ is the frequency after *N* period doublings, $f_o$ the Planck frequency (=$1/\tau_o$) and *N* the total (integer) number of doublings.

Equation for period tripling is of the same form as (36) but tripling proceeds in powers of three, as shown in Eq. (37)

$$f_N = f_o \cdot 3^{-N/3} \qquad (37)$$

If the hyperfine frequencies belong to a doubling or tripling sequence, then $N_{obs}$ should be an integer. The observed $N_{obs}$ can be solved for from Eq.'s (36) and (37) by replacing $f_N$ with the observed frequency $f_{obs}$.

The observed Hydrogen frequency yields $N_{obs}$=336.015 from Eq. (36), close to integer value 336. The Hydrogen frequency can also be obtained from Eq. (37) with $N_{obs}$=212.0015, even closer to an integer value.

The Deuterium hyperfine frequency from Eq. (37) corresponds to $N_{obs}$=216.009 period triplings, close to integer value 216. The observed Hydrogen and Deuterium frequencies have been adopted from the IAU list of astronomically important frequencies.

The observed and calculated frequencies are shown in table X together with the relative deviation of $N_{obs}$ from $N$, i.e. $Diff\% = (N_{obs}-N)/N_{obs} \cdot 100\%$.

TABLE X. Observed and calculated hyperfine frequencies and the number of doublings and triplings. Also shown is the relative difference between $N_{obs}$ and $N$.

| Object | $f_{obs}$ MHz | $f_N$ MHz | $N_{obs}$ | $N$ integer | Diff % | |
|---|---|---|---|---|---|---|
| Hydrogen | 1420.41 | 1421.20 | 212.0015 | 212 | 0.0007 | tripling |
| | | 1425.16 | 336.014 | 336 | 0.0043 | doubling |
| Deuterium | 327.39 | 328.47 | 216.009 | 216 | 0.0042 | tripling |

Table X shows that the observed number of triplings is very close to an integer number. The observed and calculated values of the frequencies differ by about 1 MHz (in tripling), the calculated frequency being higher in both cases. The calculated tripling frequencies for *N-1* and *N+1* (nearest to *N*) are more than 40% off.

### 3.17 Cosmic background temperature

The 3K cosmic background radiation corresponds to a black body, whose temperature is 2.73 K. Equation (20) yields $T$=2.76 K with $N$=320. The next number of volume doublings towards colder temperature is $N$=321, which corresponds to $T$=2.189 K.

It is interesting note that this temperature is very close to the 2.186 K $\lambda$-point of liquid helium $^4$He. This is the temperature, where helium becomes super fluid and its specific heat rises abruptly. This may be an accidental coincidence, but if it is not, then there might be a physical explanation. Since superfluidity means frictionlessness (or losslessness), then even a weak coupling to a driving force at resonant frequency (i.e. Planck sublevel) creates exchange of energy between helium and the Planck sublevels (like coupled oscillators do).

### 3.18 Cosmic ray spectrum

The most energetic particles are found in the cosmic rays producing the muon showers. The energy spectrum of incident particles is shown schematically in figure 15 as function of energy per nucleus. The vertical axis is particle flux in an arbitrary logarithmic scale. The puzzling features in the curve are the "knee" at 4.5 PeV and the "ankle" at around 6 EeV, as pointed out by R. Ehrlich (1999).

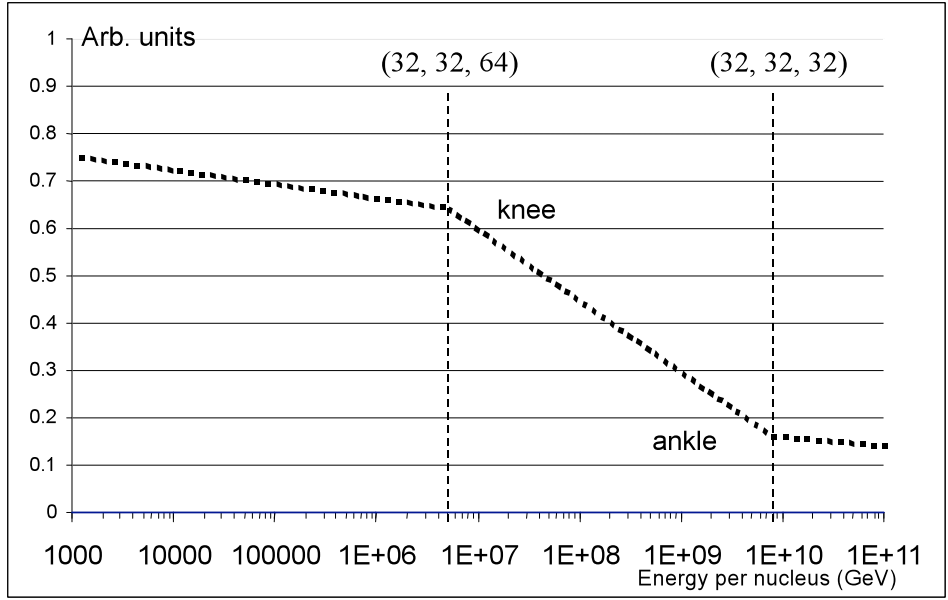

FIG. 15. The knee and ankle in the cosmic ray spectrum.

Equation (14) yields (from $E_o$) $E = 4.4$ PeV with $N = 128$. For $N = 96$ we obtain 7.1 EeV. These energies coincide with the knee and ankle energies, as shown in figure 15. The observed kinetic energies correspond to transitions from $N=125$ to $N=128$ sublevel and from $N=93$ to $N=96$ sublevels. It may me predicted that there ought to be another "knee" at $N=160$ (superstable (32, 64, 64)), corresponding to $2.7 \cdot 10^3$ GeV energy.

A review article of the present models of the knee and ankle is presented in Hörandel (2004). None of the models is based on any type of period doubling.

### 3.19 Superstability in observations

The decimal part of the perceived number of doublings tells the number of dimensions of the structure or object, because $n=N/3$ points to three dimensions and $n=N/4$ to four dimensions. This information makes it possible to break the total number of doublings (or halvings) into components in 3- and 4-dimensions. The objects in table XI are related to structures showing very high stability over time. The superstability condition $N_i=2^i$ is found with all these objects. Notation "e-p pair magnetic moment" means the sum of the absolute values of the magnetic moments of a positron and an electron.

TABLE XI. Superstable components of the perceived number ($n$) of doublings.

|  | $n$ | $N$ (3-d) | $N$ (4-d) | Components |
|---|---|---|---|---|
| Electron-positron pair (e-p) | 74.67 | 224 | 224+$l$ | (128, 64, 32, $l$) |
| e-p pair magnetic moment | 74.67 | 224 | 224+$l$ | (128, 64, 32, $l$) |
| Elementary charge squared | 9.75 |  | 39 | (1, 2, 4, 32) |
| Nucleon pair | 64.00 | 192 | 256 | (64, 64, 64, 64) |
| Nucleon pair magnetic moment | 64.00 | 192 | 256 | (64, 64, 64, 64) |
| Hydrogen spin flip | 112.00 | 336 | 448 | (128, 128, 128, 64) |
| 3K CBR | 106.67 | 320 | 320+$l$ | (128, 128, 64, $l$) |
| cosmic ray "knee" | 42.67 | 128 | 128+$l$ | (64, 32, 32, $l$) |
| cosmic ray "ankle" | 32.00 | 96 | 128 | (32, 32, 32, 32) |

The number of dimensions of the elementary electric charge squared may mean that electric interactions take place also via the fourth dimension.

## 3.20 Force ratio

If we assume that the expansion of the universe is connected to the process of volume doubling then this process should be going on at the present time, too. As no smaller electric charge is known to exist than the elementary electric charge, we must conclude that the volume doubling process is presently at halt after $N=39$ volume doublings in four dimensions. This means that an extremely stable configuration has been reached.

As comes to the Planck mass the doubling process seems to have halted at the electron-positron pair after $N=224$ volume doublings, because no lighter elementary particles are known to exist.

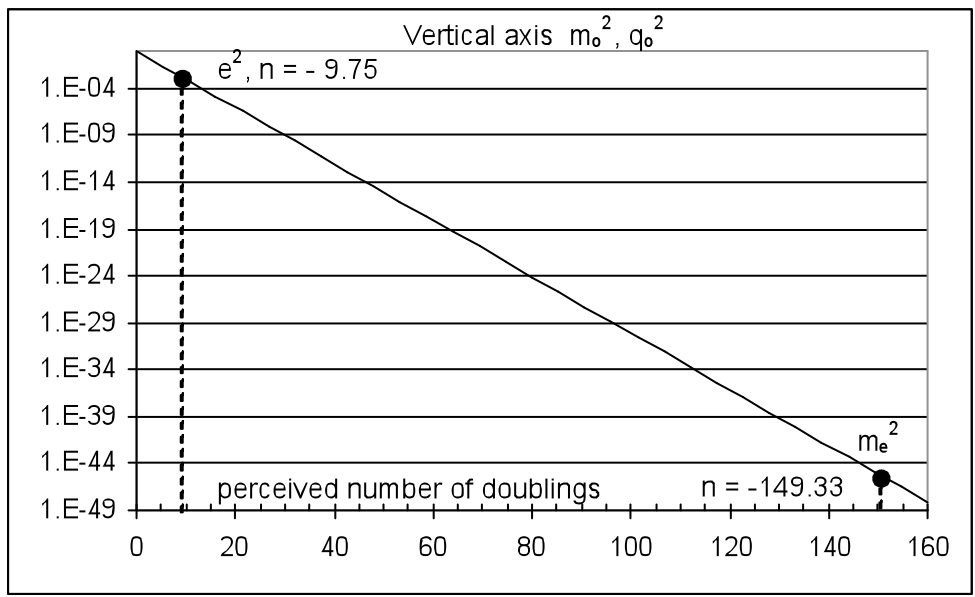

FIG. 16. Relative dilution of the Planck mass and charge squared as function of the perceived number $n$ of volume doublings in process of time. Initial value is taken as *1*. Note that $2 \cdot 74.67 = 149.33$ ($n$ of e-p pair mass squared).

The Planck mass and charge represent the same energy, because

$$Gm_o^2/l_o = q_o^2/4\pi\varepsilon_o l_o$$

This means that in the beginning the energy gradients, or the gravitational force and the electrostatic (Coulomb) force were equal.

Because the Planck charge has experienced much fewer volume doublings than the Planck mass, its strength is consequently much larger. Figure 16 shows that the volume doubling process has proceeded much farther with mass than with the electric charge. The difference between the perceived numbers of doublings is 149.33-9.75=139.58. This means that the perceived ratio of the forces (which are proportional to mass and charge squared) is $2^{(149.33-9.75)} = 10^{42}$. This ratio may be interpreted as the perceived order of magnitude ratio of the present strengths of the electric and gravitational interactions.

The strength of the so-called weak interaction is on the order of $10^{-13}$. This is close to $2^{-128/3} = 1.4 \cdot 10^{-13}$, which would relate this interaction to the superstable (32, 32, 64) sublevel.

## 4 DISCUSSION

The solution of the time-independent Schrödinger's wave equation is interpreted as the probability amplitude of finding a particle in space. Time-dependent Schrödinger's equation describes temporal evolution of the system and can be related to transitional intensities. Despite of the extreme usefulness and success of quantum mechanics, it cannot be used to solve for e.g. the intrinsic properties of the electron, like mass or charge, since they are taken as given constants in the wave equation.

Period doubling is a well established phenomenon in non-linear systems. It has been thoroughly studied both experimentally and theoretically. It is also known that nonlinear systems exhibit universal behavior which can be characterized by the Feigenbaum (1978) constants. It is period doubling that leads to (deterministic) chaos.

In this article it is shown that a differential equation derived from Kepler's law results in preferred values of orbital velocities and radii. For absolute values the Planck scale can be taken as the origin and doubling extended *ad infinitum*. The structures thus obtained are fractal by nature, since they look the same at all scales.

Solution *r(τ)* of Eq. (2) is fundamentally different from the classical Kepler's *r(τ)*, since the quantization of the Solar system, i.e. the (*r,v*)-lattice, is independent of the Sun's mass, which only determines the location of the *v=sqrt(GM/r)* hyperbola in the lattice. The orbital velocities of the planets seem to have preferred values obtainable directly from the speed of light by a 3-dimensional inverse doubling (i.e. halving) process in this case. The quantized galaxy redshifts, if interpreted as velocity, behave in the same way (Tifft 1998). This behavior of the solution of Eq. (2) may suggest a deeper connection between the model and non-linear processes in the universe.

Perturbations can be taken into account by adding attenuation (or amplification) in Eq. (2) such that the constant of attenuation *b/τ* is inversely proportional to period.

$$\frac{d^2r}{d\tau^2} = -\frac{a}{\tau^2}r - \frac{b}{\tau}\frac{dr}{d\tau} \qquad (38)$$

Several other experimentally determined properties of invariants come into the realm of the solution of Eq. (2). If the circumference *2πr* is regarded as wavelength *λ*, then also energies *E=hc/λ* become quantized. If the Planck length is taken as the initial *λ* (i.e. *E*=Planck energy), then a good fit with the properties of the basic elementary particles is obtained. Magnetic moments can be likewise calculated if the elementary charge is given. The electric energy seems to follow 4-dimensional doubling (or halving) suggesting that the electric interactions may require more than three dimensions.

According to Eq. (16) the model allows particles to take on other values of the electric charge than the elementary charge. This possibility should be taken into account in analyzing experimental particle mass spectra in view of this model. Electrons and protons have very long lifetimes, which mean that the associated structures are extremely stable.

One possible scenario for the superstability is shown in chapter 2.8, where a special subset of possible doublings is created. Table XI shows that the superstability condition is rather general. The real physical processes behind the superstability remains unsolved for the time being.

The rest energies of the electron and the positron are found in the direct sequence of energy sublevels obtained by the doubling process from the Planck energy. The nucleons seem to be composite particles, because their rest energy is obtained from a sum energy level.

The magnetic moments of the electron and proton seem to originate from two different geometries. The unit current loop defining the electron's magnetic moment is Bohr-type, i.e. the circumference of the loop is the Planck length whereas the geometry of the proton's unit magnetic moment is potential well ground state type (loop diameter is half of the Planck length).

Spin was originally invented to explain some features of the atomic spectra. It is also used to explain the existence and magnitude of magnetic moments of particles together with the associated experimental gyromagnetic ratio, which depends on the internal (largely unknown) structure of particles. Difficulties with the relation of spin and the internal structure are manifested in the notorious proton spin problem. In this model it is simply the classical magnetic moment that determines the magnetic moments of the particles dealt with.

The analysis of the electron magnetic moment anomaly in chapter 3.12 showed that according to this model the anomaly is negative and much smaller than the classical anomaly.

Nonlinear dynamical systems exhibit doubling, tripling, quarupling and so forth of the fundamental period. The doubled periods double, tripled periods triple again and so forth and corresponding frequencies appear. The period tripling process may be responsible for the hyperfine structures of Hydrogen and Deuterium, as shown in table X.

The superstability condition applies to the electron-positron pair, not the electron or positron alone. This is interesting in the sense that it is the electron-positron pair that is stable. This implies that the electron and positron always co-exist forming a joint energy-object and being somehow connected. The pair can be spatially separated in the 3-dimensional world without altering their invariant properties. Freedom in the 3-dimensional space means that the connection may be via the fourth dimension.

The nucleons can be thought to originate from a superstable (64, 64, 64, 64) $E_{os}$ sublevel. This level splits into two nucleons the same way as the electron and positron are borne. Table VIII shows that 1.021 MeV of additional energy is needed in order to obtain the measured rest energy of the proton. The corresponding two 511 keV structures are (33, 65, 129), i.e. the electron or positron structure, but only one of the levels becomes charged (positively in this case) and can be thought to produce the (65)-level magnetic moment of the proton. The neutron seems to host both charges together with an additional 1.28 MeV energy, which is the $N=223$ ($E_o$) structure. The length corresponding to level (65) is $2^{65}l_o = 1.5 \cdot 10^{-15}$ m, which is of the order of the nucleon "size".

The elementary particle processes seem to be

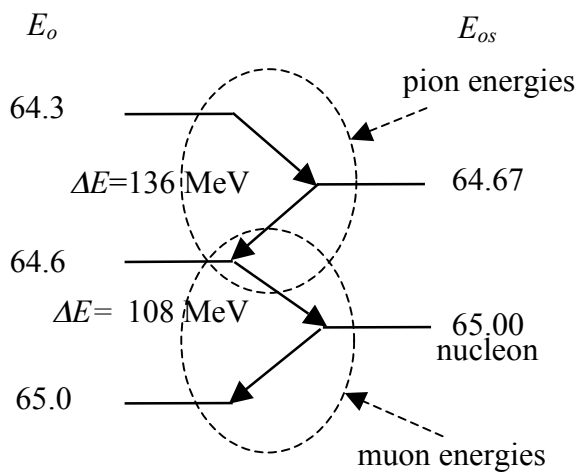

FIG. 17. Pion and muon energies appear in transitions around the nucleon level.

governed by the 3-d or 4-d structural changes appearing as changes in the number of doublings (as e.g. in pion decay into a muon, table VII)

It is known that pions are profusely produced in proton collisions. Figure 17 shows that the energy level differences around the nucleon (65)-level correspond to the pion and muon energies. These energies may be transferred to the pion and muon energy levels by resonant energy transfer.

The distances $r_n$ (in A.U.) of the planets from the Sun obey the so called Titius-Bode rule, found experimentally and not valid beyond Uranus. One form of the rule is $r_n=n+0.4$ in astronomical units with *n=0, 0.3, 0.6, 1.2, 2.4 …* (doubling *n*, *n=0* for Mercury). The difference between the Titius-Bode rule and the model presented in this article is that the quantized Keplerian system gives absolute values for the allowed orbits (and orbital velocities in addition).

The cosmic background radiation (CBR) is detected as microwave photons coming from all over the sky. In quantum systems electromagnetic quanta are emitted during transitions from a higher energy level to a lower one. The CBR temperature converted into energy of the photons corresponds to the superstable $N=320$ sublevel energy. This means that the transition occurs from $N=319$ level to $N=320$ level, whence the photon energy appears as $N=320$ level energy, which is the energy difference of these levels. Corresponding reasoning applies to the 21 cm wavelength of the Hydrogen spin-flip.

W.G. Tifft (1996) has shown that the redshift periods of galaxies may change from one period to the next (redder) period in a few years. As the thicknesses and diameters of galaxies are typically 10,000-100,000 light years, no signal traveling at the speed of light would reach the whole galaxy within years. This implies that a galaxy is some kind of a superstructure and the change is due to the doubling process concerning the whole structure simultaneously. According to this model the change must take place in the fourth dimension.

The experimental data suggests that the observer seems to perceive the cube root or the fourth root of a volume, or the geometric mean of the edge lengths of the corresponding three and four dimensional geometric structures. This situation may be due to the way we carry out measurements, which produce scalar values.